\documentclass{aa}
\usepackage{graphicx}
\newcommand{\xmm}{\textit{XMM-Newton}\xspace}
\newcommand{\chandra}{\textit{Chandra}\xspace}

\usepackage{natbib,twoopt}
\usepackage{xcolor}
\usepackage[breaklinks=true, colorlinks=true, linkcolor=blue, citecolor=blue]{hyperref} 
\bibpunct{(}{)}{;}{a}{}{,}             
\makeatletter
\newcommandtwoopt{\citeads}[3][][]{\href{http://adsabs.harvard.edu/abs/#3}%
    {\def\hyper@linkstart##1##2{}%
     \let\hyper@linkend\@empty\citealp[#1][#2]{#3}}}
  \newcommandtwoopt{\citepads}[3][][]{\href{http://adsabs.harvard.edu/abs/#3}%
    {\def\hyper@linkstart##1##2{}%
     \let\hyper@linkend\@empty\citep[#1][#2]{#3}}}
  \newcommandtwoopt{\citetads}[3][][]{\href{http://adsabs.harvard.edu/abs/#3}%
    {\def\hyper@linkstart##1##2{}%
     \let\hyper@linkend\@empty\citet[#1][#2]{#3}}}
  \newcommandtwoopt{\citeyearads}[3][][]%
    {\href{http://adsabs.harvard.edu/abs/#3}
    {\def\hyper@linkstart##1##2{}%
     \let\hyper@linkend\@empty\citeyear[#1][#2]{#3}}}
\makeatother
\usepackage{multirow}
\usepackage{comment}
\usepackage{siunitx}  
\usepackage{upgreek}
\usepackage{txfonts}
\usepackage{subfig}
\usepackage{sidecap}
\sidecaptionvpos{figure}{c}
\begin{document} 

   \title{The Quiescent Sloshing Core of Abell 496 with XRISM}
   \subtitle{}
   \titlerunning{The Quiescent Sloshing Core of Abell 496 with XRISM}

   \author{Angie Veronica\inst{1}
          \and
          Thomas H. Reiprich\inst{1}
          \and
          Naomi Ota\inst{1,2}
          \and
          Jakob Dietl\inst{1}
          \and
          Frederick Groth\inst{3}
          \and
          Klaus Dolag\inst{3,4}
          \and
          Efrain Gattuzz\inst{5}
          \and
          Florian Pacaud\inst{1}
          \and
          Elke Roediger\inst{6}
          \and
          Jeremy S. Sanders\inst{5}
          \and
          Benjamin Seidel\inst{3}
          \and
          Yuanyuan Zhao\inst{1}
          }

   \institute{Argelander-Institut f\"ur Astronomie (AIfA), Universit\"at Bonn, Auf dem H\"ugel 71, 53121 Bonn, Germany\\
        \email{averonica@astro.uni-bonn.de}
        \and
        Nara Women's University, Kitauoyanishi-machi, Nara, 630-8506, Japan
        \and
        Universitäts-Sternwarte, Fakultät für Physik, Ludwig-Maximilians-Universität München, Scheinerstr. 1, 81679 München, Germany
        \and
        Max-Planck-Institut für Astrophysik, Karl-Schwarzschild-Straße 1, 85741 Garching, Germany
        \and
        Max-Planck-Institut f\"ur extraterrestrische Physik, Gießenbachstraße 1, 85748 Garching, Germany
        \and
        E.A Milne Centre for Astrophysics, University of Hull, Cottingham Road, Hull HU6 7RX, UK
        }
   \date{Received DD MM YYYY / Accepted DD MM YYYY}

 
  \abstract
   {Gas motions provide insight into the dynamical history and physical processes within galaxy clusters. The nearby, X-ray bright, strong cool-core cluster Abell 496 (A496) is a unique system. Strong gas sloshing features have been observed, and detailed simulations have predicted that these were induced by a previous north-south minor merger. Consistent with this scenario, SRG/eROSITA All-Sky Survey data confirm an emission enhancement in the northwest direction.}
   {We investigate the kinematics of the intracluster medium (ICM) in the core of A496 using high-resolution data from the Resolve micro-calorimeter on board XRISM.}
   {The Resolve A496 observation enables us to directly probe the ICM motions in the cluster core, measuring line-of-sight (LOS) bulk and turbulent velocities via line shifts and broadening of characteristic lines such as Fe\,\textsc{xxv}\,He$\alpha$ and Fe\,\textsc{xxvi}\,Ly$\alpha$. We compared our measurement with other Resolve cluster core measurements and further compared our results with simulations and multiwavelength observations.}
   {From an optical redshift analysis, we found that the BCG is at rest with respect to the systemic velocity of the cluster. Despite multiple previously detected cold fronts and harboring a weak central radio source, Resolve observation shows that the core of A496 is dynamically quiescent. The ICM is moving with respect to the brightest cluster galaxy with a LOS bulk velocity of $v_{\rm bulk}=-69_{-20}^{+25}\,\mathrm{km\,s}^{-1}$. We measured a turbulent velocity of $\sigma_{\rm v}=78_{-16}^{+18}\,\mathrm{km\,s}^{-1}$, the lowest value reported by the instrument on a cluster core to date. This value is also in good agreement with the velocity dispersion of the H$\alpha$ filament in the core, which may indicate condensation of ICM in the wake of the radio bubble. Assuming isotropic turbulence, the ICM turbulent velocity corresponds to a subsonic 3D Mach number of $0.15_{-0.03}^{+0.04}$ and a non-thermal pressure fraction of $1.2_{-0.5}^{+0.6}\,\%$. The mechanical AGN feedback from the recent activity of the central radio source is estimated to contribute about $7$--$9\,\%$ to the ICM heating. The 1D LOS bulk velocity from the Simulating the LOcal Web (SLOW) constrained Universe simulation is consistent with the measured value, suggesting that AGN feedback has a negligible contribution. The A496 SLOW turbulent velocity, as in other reported Resolve$-$simulation comparisons, is higher, but remains within $1.5\sigma$ uncertainty.
   }
   {A496 may represent one of the most quiescent sloshing cores observed so far.}

   \keywords{Galaxies: clusters: individual: Abell 496 -- X-rays: galaxies: clusters -- Galaxies: clusters: intracluster medium}

   \maketitle
\section{Introduction}\label{sec:intro}
Galaxy clusters form and grow both through major mergers and continuous accretion of lower-mass systems along cosmic web filaments \citep[e.g.,][]{Kravtsov_2012}. These events leave imprints at various scales in the hot plasma ($10^7\,{\rm K}$ to $10^8\,{\rm K}$), the intracluster medium (ICM), that permeates between the member galaxies. Major mergers produce distinct large-scale disturbances in the ICM, such as shocks, bulk flows, and turbulence \citep{Markevitch_2007}.
On smaller scales, gas motions are driven by localized activities, such as sloshing, AGN feedback, and galaxy motion. In the centers of clusters, active galactic nuclei (AGNs) harbored by the brightest cluster galaxies (BCGs) redistribute their energy through low-density, rising bubbles inflated by relativistic AGN jets \citep[for reviews see, e.g.,][]{Fabian_2012, Hlavacek_2022}. The so-called AGN feedback has been known to contribute to the heating of the ICM in cool-core clusters \citep[e.g.,][]{Mittal_2009, Pasini_2022, Veronica_2026}, which is also expected to cause bulk and turbulent motions.
\par
Studying gas motions is non-trivial and has previously been challenging due to a modest spectral resolution of the CCD-based X-ray instruments, which is of the order of $\sim\!150\,{\rm eV}$. The Hitomi satellite was launched in 2016 \citep{Takahashi_2016} and carried a non-dispersive X-ray microcalorimeter with an unprecedented spectral resolution of $4.9\,{\rm eV}$. Despite its short mission life, the instrument successfully returned precise measurements of gas motions in the center of the Perseus cluster \citep{Hitomi_2016, Hitomi_2018}. Hitomi is then succeeded by the X-ray Imaging and Spectroscopy Mission (XRISM), which was launched in September 2023 \citep{Tashiro_2025}. XRISM carries two science instruments: the Resolve soft X-ray spectrometer \citep{Ishisaki_2022} and the Xtend soft X-ray imager \citep{Noda_2025}. Resolve is an X-ray microcalorimeter that consists of a $6\times6$ pixel array. It has a field-of-view (FoV) of $3.1'\times3.1'$ and an energy resolution of $4.5\,{\rm eV}$ FWHM at $6.0\,{\rm keV}$. The Xtend instrument is equipped with a CCD camera with a wide FoV of $38.5'\times38.5'$ with a moderate energy resolution of $\sim\!180\,{\rm eV}$ FWHM at $6.0\,{\rm keV}$. The XRISM/Resolve observations enable us to directly probe ICM motions, namely the line-of-sight (LOS) bulk velocity and turbulent velocity, through measurements of line shifts and broadening, respectively. Resolve measurements of numerous clusters have been reported, including some known cool core (CC) clusters, e.g., A2029 \citep{XRISM_A2029, XRISM_A2029_Sarkar}, Ophiuchus \citep{Fujita_2025}, Centaurus \citep{XRISM_Cent}; and dynamically more active, e.g., Virgo \citep{XRISM_Virgo} and A3395S \citep{Ota_2026}. The former sample exhibits a systematically lower velocity dispersion ($<170\,{\rm km\,s^{-1}}$) than the latter ($\sim\!260\,{\rm km\,s^{-1}}$).
\par
\begin{figure}
\centering
    \includegraphics[width=\columnwidth]{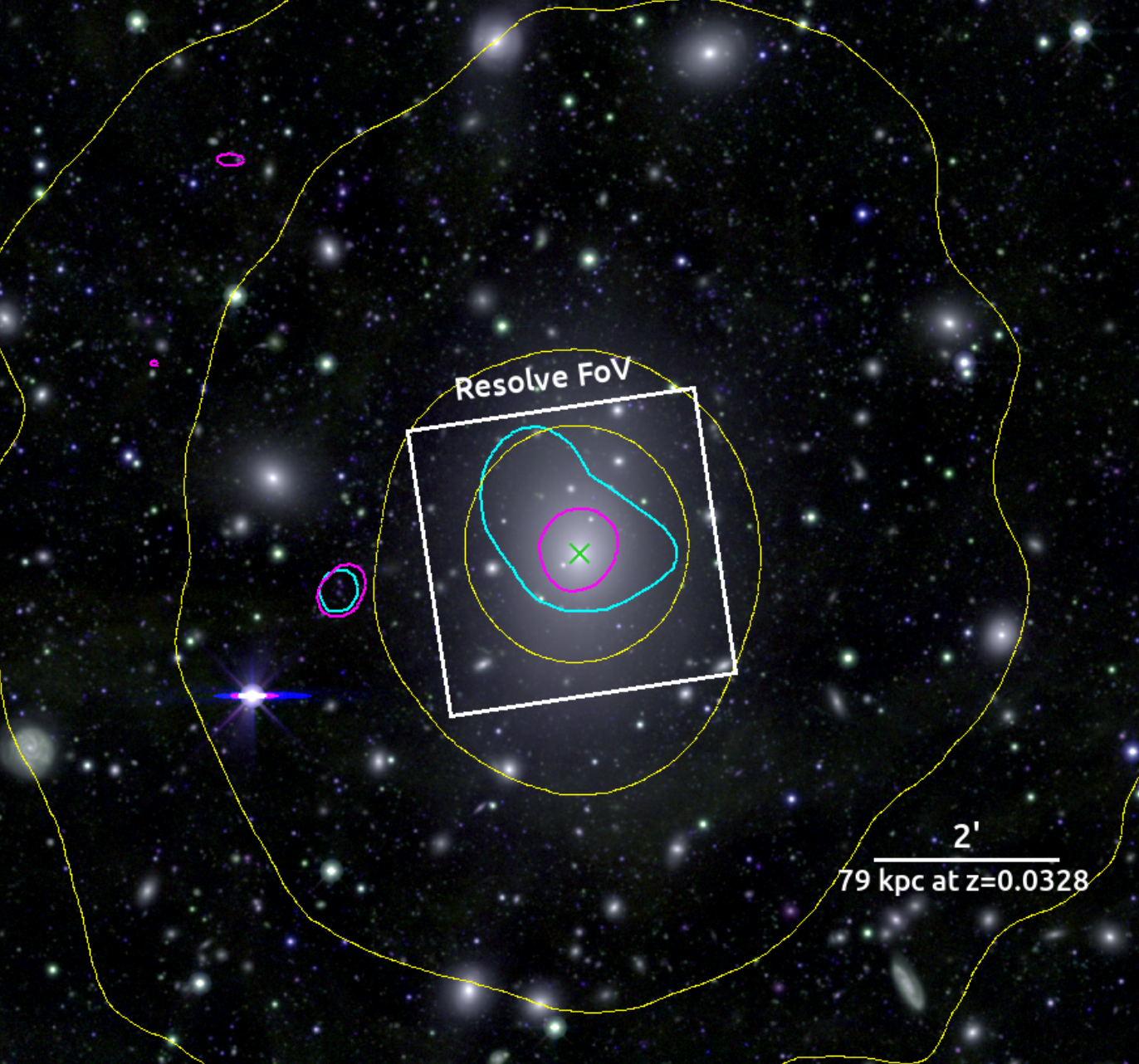}
\caption{Composite image of A496. The background image is the r, g, z DESI Legacy Survey DR10 in red, green, and blue channels, respectively.
Overlaid contours are the GMRT $150\,\mathrm{MHz}$ emission at $0.03\,\mathrm{Jy\,beam^{-1}}$ (cyan), RACS $887.5\,\mathrm{MHz}$ radio emission at $0.00225\,\mathrm{Jy\,beam^{-1}}$ (magenta), and eRASS:5 X-ray emission at $0.001,\,0.0004,\,0.0001,\,0.00005\,\mathrm{cts\,s^{-1}}$ (yellow). The white box indicates the Resolve FoV and the green cross marks the BCG position.}
\label{fig:rgbopt}
\end{figure}

In this work, the results of the XRISM observation of the galaxy cluster Abell 496 (A496) are presented. A496 is an X-ray bright and nearby galaxy cluster ($z=0.0328$), identified as a strong cool-core cluster with a central cooling time of approximately $0.5\,{\rm Gyr}$ with a high core metallicity of $0.66Z_\odot$ \citep{Hudson_2010}. Fig.~\ref{fig:rgbopt} shows a composite image of the central region of A496. The background image is the DESI Legacy Survey DR10 g, r, z image in the red, green, and blue channels, respectively. The yellow contour shows the X-ray emission from the five combined SRG/eROSITA All-Sky Survey \citep{Predehl_2021, Sunyaev_2021, Merloni_2024}. Cyan and magenta contours mark the extended radio emission from TIFR GMRT Sky Survey \citep[TGSS;][]{Intema_2017}\footnote{\url{https://tgssadr.strw.leidenuniv.nl/doku.php?id=start}} at $150\,\mathrm{MHz}$ emission and the Rapid ASKAP Continuum Survey \citep[RACS;][]{RACS}\footnote{\url{https://www.atnf.csiro.au/research/skyviewer/}} at $887.5\,\mathrm{MHz}$ emission above $3\sigma$ of their rms, which are $0.03\,\mathrm{Jy\,beam^{-1}}$ and $0.00225\,\mathrm{Jy\,beam^{-1}}$, respectively. The FWHM of the radio beam is $25\times25''$ for TGSS and $15\times15''$ for RACS. The different sizes of the radio structures suggest radio emission from central AGN activities at different epoch, with TGSS showing emission from an older relativistic electron population and RACS from a more recent activity. The green cross indicates the position of the brightest cluster galaxy (BCG), and the white box is the Resolve FoV of the A496 observation.
\par
\begin{figure*}[!ht]
\centering
    \includegraphics[width=0.98\textwidth]{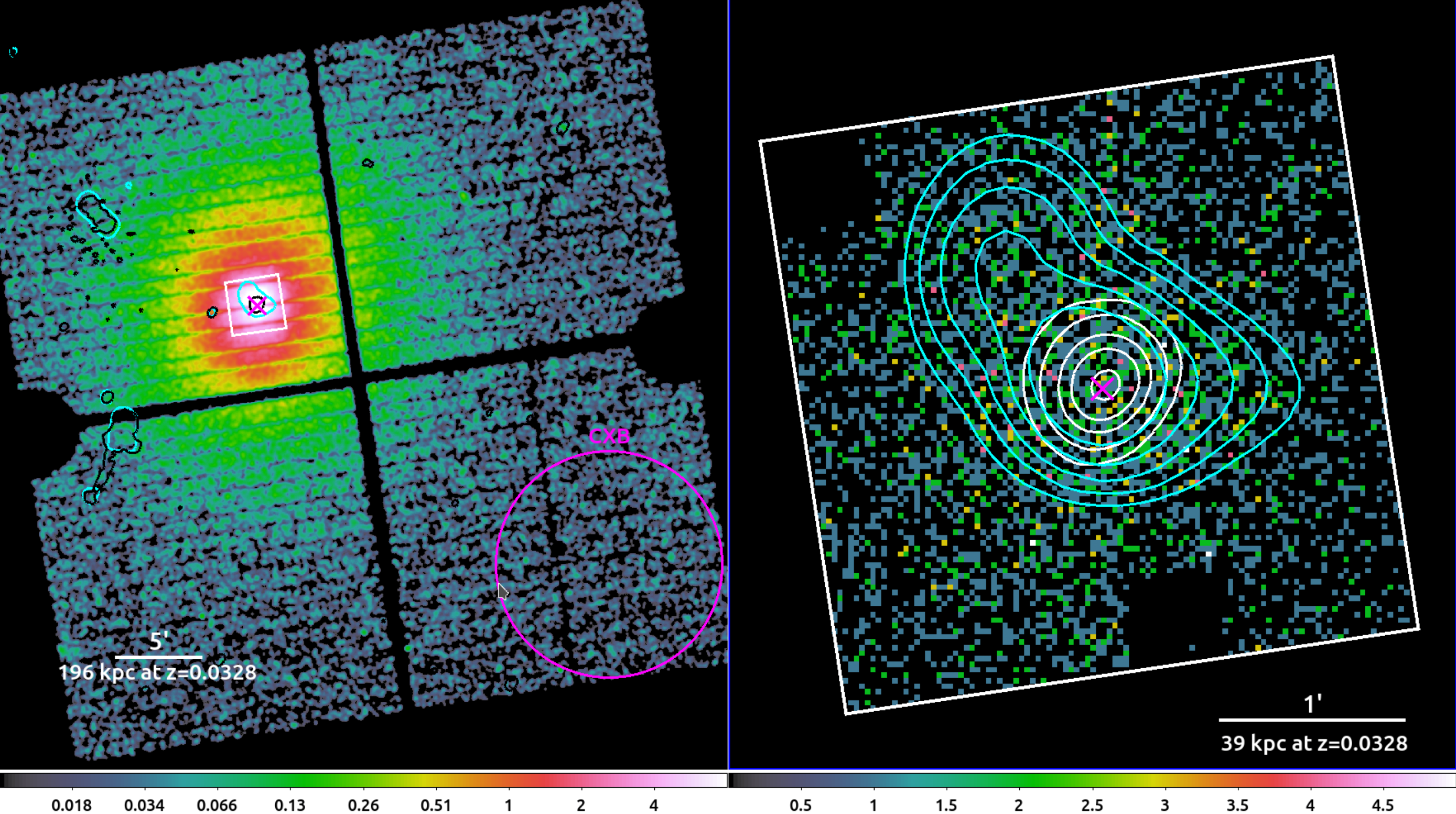}
\caption{XRISM observation of A496. In both panels, white box indicates the Resolve FoV and magenta circle shows the CXB region. The magenta cross marks the position of the BCG. The cyan and black contours are the TGSS at $150\,\mathrm{MHz}$ and RACS at $887.5\,\mathrm{MHz}$ radio contours, respectively. \textit{Left}: Xtend photon image in the $0.3-10.0\,{\rm keV}$ band. The radio contours are at $3\sigma$ RMS; $0.03\,\mathrm{Jy\,beam^{-1}}$ for GMRT and $0.00225\,\mathrm{Jy\,beam^{-1}}$ for RACS. \textit{Right}: Resolve photon image in the $2.0-10.0\,{\rm keV}$ band. The TGSS contour (cyan) is at 0.03, 0.04, 0.09, 0.4, and $2.0\,\mathrm{Jy\,beam^{-1}}$, while the RACS contour (white) is at 0.00225, 0.01, 0.06, 0.36, $2.0\,\mathrm{Jy\,beam^{-1}}$.}
\label{fig:xrism}
\end{figure*}

Over decades, A496 has been the subject of targeted and sampled studies in various wavelengths. In X-ray, the cluster appears relaxed, with the X-ray peak coinciding well with the position of the central dominant galaxy (cD), which is also its BCG \citep{Dupke_2007}. In general, the X-ray morphology of the cluster is elongated in the northwest direction (see the yellow contour in Fig.~\ref{fig:rgbopt} and Xtend image in the left panel of Fig.~\ref{fig:xrism}). However, detailed investigations of the core using \xmm and \chandra observations reveal a spiral pattern \citep{Lagana_2010} and several cold fronts associated with sloshing \citep[e.g.,][]{Dupke_2007, Ghizzardi_2014}. Dedicated hydrodynamical simulations show that these observed features are well reproduced by an off-axis minor merger, which induced a sloshing motion of the core \citep{Roediger_2012}. A496 hosts a central radio galaxy detected across multiple frequencies \citep{Hogan_2015, Hogan_2015hi, Ubertosi_2024}. To investigate the interplay between the central radio galaxy of A496 and its immediate atmosphere, \cite{Ubertosi_2024} utilized multiple deep radio data including data from the GMRT at 150, 330, and $617\,{\rm MHz}$, the VLA at 1.4 and $4.8\,{\rm GHz}$, and the VLA Low Band Ionosphere and Transient Experiment (VLITE) at $340\,{\rm MHz}$, as well as \chandra X-ray and Very Large Telescope (VLT) MUSE observations. From multi-frequency radio images, they identified central AGN activities from three different episodes indicated by the spectral index steepness of the emission (synchroton aging), going from flat at higher frequencies on the sub-kpc scales (recent), steep on the $\sim\!20\,{\rm kpc}$ scales (older), and ultrasteep on the $\sim\!50-100\,{\rm kpc}$ scales (oldest).
\par
The structure of this paper is as follows: In Sect.~\ref{sec:drsteps}, the data reduction steps and the analysis strategy of the A496 XRISM data are described. In Sect.~\ref{sec:results}, the results are presented, and they are discussed in Sect.~\ref{sec:discussion}. We present the conclusion and summary of this work in Sect.~\ref{sec:conclude}. Unless stated otherwise, all uncertainties are at the 68.3\% confidence interval. The assumed cosmology in this work is a flat $\Lambda$CDM cosmology, where the Hubble constant is $H_0=70~\mathrm{km~s^{-1}~Mpc^{-1}}$, $\Omega_{\rm m} = 0.3$, and $\Omega_\Lambda=0.7$.
At the redshift of the BCG of the A496 cluster, $z_{\rm BCG}=0.0328$ \citep{Wegner_1999}, $1''$ corresponds to $0.655\,{\rm kpc}$.

\begin{table}
\centering
\caption{Information on the Abell 496 cluster used in the present work.}
\begin{tabular}{c c}
\hline
\hline
R.A., Dec. [J2000] & $68.4082^{\circ}, -13.2611^{\circ~\dagger}$ \\
$z_{\rm BCG}$ & $0.03281\pm0.00004^a$\\
$M_{500}$ & $(2.76\pm0.11)\times10^{14}M_\odot^b$ \\
$R_{500}$ & $1376''^b$ \\ 
$R_{200}^c$ & $2175''$\\[2pt]  
\hline
\multicolumn{2}{l}{\footnotesize $^\dagger$ \xmm X-ray peak, $^a$\cite{Wegner_1999},}\\
\multicolumn{2}{l}{\footnotesize $^b$\cite{RB_2002}}\\
\multicolumn{2}{l}{\footnotesize $^cR_{200}\approx R_{500}/0.65$ \citep{Reiprich_2013}}\\
\hline
\hline
\end{tabular}
\label{tab:A496info}
\end{table}

\section{XRISM A496 observation}

The core of the A496 cluster was observed by XRISM on 24 March 2025 as a part of the AO1 (OBSID: 201122010, P.I.: T. H. Reiprich). The pointed observation was carried out using both XRISM instruments: the Resolve high-resolution microcalorimeter and the Xtend soft X-ray imager. The information on the XRISM A496 observation is listed in Table~\ref{tab:obslog}. In this work, we utilized the Resolve data to estimate the ICM properties in the core of A496, while we assessed the sky background component using the Xtend data. The data screening and reduction steps were performed following the prescription outlined in The XRISM Data Reduction Guide and \cite{Ota_2026} for the analyses.

\begin{table*}[!ht]
\centering
\caption{XRISM observation of A496.}
\resizebox{0.9\textwidth}{!}
{\begin{tabular}{c c c c c c c}
\hline
\hline
Object & OBSID & Coordinates$^a$ & DATE-OBS & Exposure$^b$ & Fe\,\textsc{xxv}\,He$\alpha^c$ & Fe\,\textsc{xxvi}\,Ly$\alpha^c$ \\
\hline
ABELL496\_CENTER & 201122010 & $68.4088^\circ, -13.2616^\circ$ & 2025-03-24 & 23.1 & 376 & 33 \\
\hline
\multicolumn{7}{l}{\footnotesize $^a$Pointing coordinates, $^b$exposure time after screening in ks, $^c$ Iron line counts without the continuum component.}\\
\hline
\hline
\end{tabular}}
\label{tab:obslog}
\end{table*}

\subsection{Data reduction}\label{sec:drsteps}
Data reduction and analysis of Resolve and Xtend data were performed using the XRISM tasks implemented in \texttt{HEASoft~v6.35} and using the calibration database (CALDB) version 20250915. They are summarized below.
\par
For the Resolve data, firstly, the event file was reprocessed using the latest CALDB. Frame events were then screened by applying PI and rise time cuts, as well as \texttt{STATUS[4]}. The clean exposure time is $23.1\,\mathrm{ks}$. A lightcurve in the $2.0-10.0\,{\rm keV}$ energy band was inspected and showed no temporal variation throughout the observation. For the spectral analysis, we only included High-primary events (Hp; \texttt{ITYPE==0}) that achieve the highest Resolve energy resolution. We excluded pixel 12 (the calibration pixel) and pixel 27, which shows different gain variation characteristics from the other pixels.
\par
For the Xtend data, after reprocessing, we removed the anomalous pixels in the Xtend field-of-view (FoV) using the \texttt{xtdpixclip} task. The task was run in two steps: in the \texttt{HISTO} mode to build a histogram of the supplied region file, then in the \texttt{APPLY} mode, when the threshold cut for the provided region was applied. A clean event file was generated by the task when \texttt{mkclean=yes} parameter was given. The fully cleaned Xtend photon image in the $0.3-10.0\,{\rm keV}$ band and the Resolve photon image in the $2.0-10.0\,{\rm keV}$ band are shown in Fig.~\ref{fig:xrism}. In the Resolve photon image the counts are randomized in location within the detector pixels.

\subsection{Spectral analysis}\label{sec:spectro}

\begin{figure*}
\centering
    \includegraphics[width=0.49\textwidth, trim=1.5cm 1.25cm 3.0cm 2.0cm, clip]{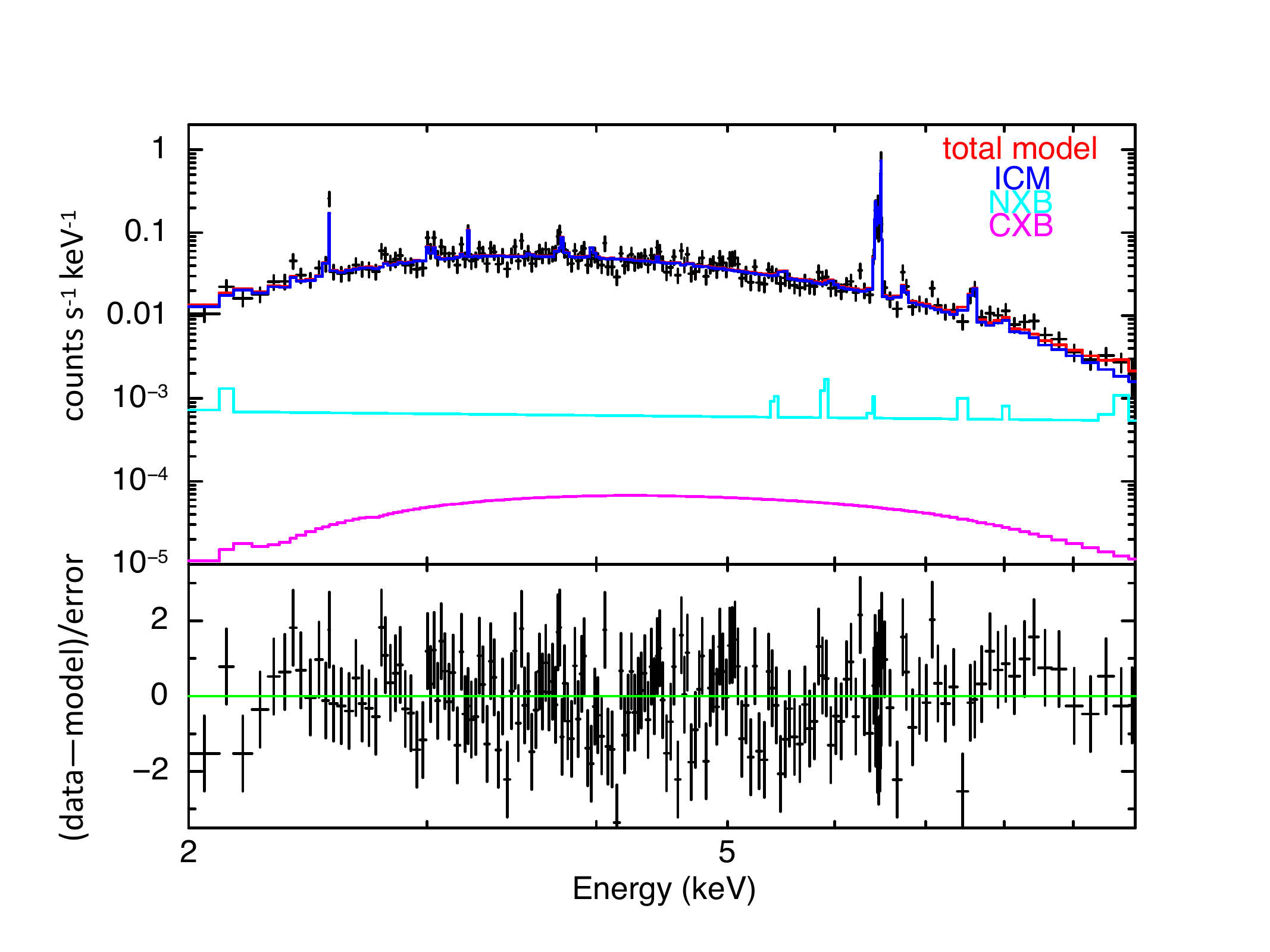}
    \includegraphics[width=0.49\textwidth, trim=1.5cm 1.25cm 3.0cm 2.0cm, clip]{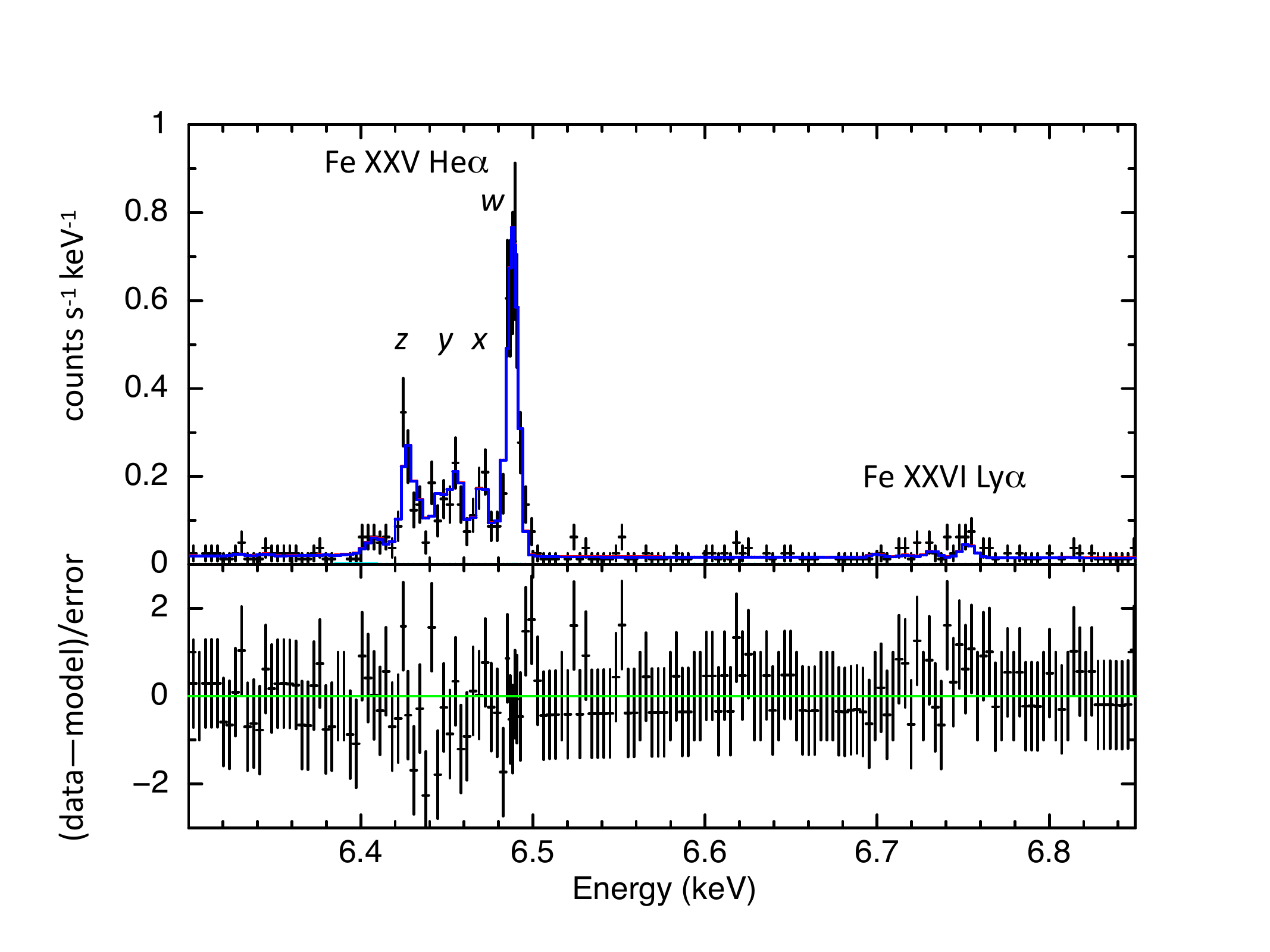}
\caption{XRISM/Resolve spectrum of the core of A496. In both panels, the black crosses are the data points from the observation. The total model (red line) is the sum of the absorbed cluster emission represented by \texttt{tbabs}$\times$\texttt{bapec} (blue), the NXB extracted from the night-Earth observation (cyan), and the CXB from the Xtend analysis (magenta). \textit{Left}: The spectrum is shown in broad $2.0-10.0\,{\rm keV}$ band and grouped to achieve at least $5\sigma$ per bin for visualization purposes. \textit{Right}: Zoom-in to show the Fe\,\textsc{xxv}\,He$\alpha$ and Fe\,\textsc{xxvi}\,Ly$\alpha$ complexes. The spectrum is grouped in sets of seven bins ($3.5\,{\rm eV}$).}
\label{fig:rsl_spec}
\end{figure*}

All spectral fittings were performed with the X-ray spectral fitting package \citep[\texttt{XSPEC};][]{XSPEC} version 12.15.0 and the atomic database \texttt{atomDB~v3.0.9} \citep{Foster_2012_atomdb}. For estimating the errors of the parameters, the C-statistic \citep{Cash_1979} was adopted, and the proto-solar abundance table from \cite{Lodders_2009} was used.
\par
The Resolve response matrix file (RMF) was generated using \texttt{rslmkrmf} task, specifying \texttt{L} size. The ancillary response file (ARF) was constructed by the \texttt{xaarfgen} task, using as an input the fully-corrected image of the five combined SRG/eROSITA All-Sky Surveys (eRASS:5) in the broad energy band of $0.3-10.0\,\mathrm{keV}$ and with a radius of $3'$. The eRASS:5 data were processed using the eROSITA Science Analysis Software System \citep[\texttt{eSASS};][]{Brunner_2022} version~240410.0.3, and the data reduction and image correction steps follow the description in previous eROSITA works, e.g., \cite{Veronica_2025}.
\par
We utilized the Xtend data to estimate the contribution of the cosmic X-ray background (CXB). We extracted a spectrum from a circular region farthest from the cluster within the FoV (magenta circle in the left panel of Fig.~\ref{fig:xrism}). The RMF was produced with the \texttt{xtdrmf} task and the ARF with the \texttt{xaarfgen} task, assuming a uniform sky with the \texttt{FLATCIRCLE} mode with a flat radius of $15'$. The CXB spectrum was fitted with the typical CXB model, comprising a thermal emission to account for the Local Hot Bubble (LHB), an absorbed thermal emission for the Milky Way halo (MWH), and an absorbed power law for unresolved sources. Except for normalizations, the parameters of these components were frozen to the values listed in Table~3 in \cite{Veronica_2024} during the fitting. The best-fit values from Xtend were then utilized to model the CXB in the Resolve spectral analysis. We note that using the best-fit normalization values obtained from eRASS:5 spectra of the same region yields consistent results.
\par
For both XRISM instruments, the extractions and treatments of the non--X-ray Background (NXB) for the A496 observation were carried out by following the description in the XRISM Non-X-ray Background Database and Tools\footnote{\url{https://heasarc.gsfc.nasa.gov/docs/xrism/analysis/nxb/index.html}}. The NXB spectra were derived from the night-Earth database, and we applied the same screening as the source data. The NXB spectrum was extracted using the \texttt{rslnxbgen} task for Resolve and the \texttt{xtdnxbgen} task for Xtend. Each spectrum was then fitted using the respective spectral model provided by the XRISM team, and the resulting best-fit model was applied to the corresponding science analyses.
\par
The complete model of Resolve A496 spectral fitting was the sum of the absorbed cluster emission, the CXB, and the NXB. The cluster component was modeled using \texttt{bapec} \citep{Smith_2001} multiplied by the absorption along the line of sight by the Galactic column density parametrized by \texttt{tbabs} \citep{Wilms_2000}. The Galactic column density was fixed to the average total hydrogen column density in the corresponding analyzed region, $N_\mathrm{H,tot}$, which was calculated by adding the HI4PI all-sky Galactic neutral atomic hydrogen map \citep{HI4PI_2016} to the $N_{\mathrm{H}_2}$ map calculated using the method described in \cite{Willingale_2013} through the Swift Galactic column density of Hydrogen tool\footnote{\url{https://www.swift.ac.uk/analysis/nhtot/}}. The $N_\mathrm{H,tot}$ for the Resolve FoV and the Xtend CXB region are $6.5\times10^{20}\,\mathrm{cm}^{-2}$ and $6.9\times10^{20}\,\mathrm{cm}^{-2}$, respectively. All the \texttt{bapec} parameters, namely, temperature ($k_\mathrm{B}T$), metallicity ($Z$), redshift ($z$), velocity broadening ($\sigma_{\rm v}$), and normalization, were left to vary. The CXB component was fixed to the resulting values from Xtend CXB analysis and folded with Resolve flat ARF, while the NXB component was fixed to the results of the previously mentioned NXB analysis.

\subsubsection{Systematic uncertainties}\label{sec:systematics}
To investigate any systematic effects on the robustness of the estimated redshift (hence, LOS bulk velocity) and turbulent velocity, several additional analyses were performed. First, the impact of spectrum grouping was examined by additionally fitting binned spectra with a minimum of 1 count and 2 counts per bin. From all three fittings (unbinned, 1 count binned, and 2 counts binned), all parameters are in good agreement with each other within their statistical uncertainties.
\par
Then, the impact of the assumed energy band was investigated by also performing spectral fitting in the restricted energy band to mainly include the Fe\,\textsc{xxv}\,He$\alpha$ and Fe\,\textsc{xxvi}\,Ly$\alpha$ line complexes, namely, $5.5-7.0\,\mathrm{keV}$. Fitting in this narrow energy band results in a small increase in the statistical errors, that is, about $13\%$ and $9\%$ for redshift and turbulent velocity, respectively. The increase in uncertainties in these two parameters might be propagated from the reduced constraining power of other relevant parameters in this narrow band. For instance, the statistical error in the temperature increases by almost a factor of three. Alternatively, other line complexes in the broad energy band might contribute to some constraining power. The best-fit values obtained for these parameters in the broad $2.0-10.0\,\mathrm{keV}$ and narrow energy bands are consistent within their $1\sigma$ significances, with a relative difference of much less than $0.1\%$ for the redshift and about $10\%$ for the turbulent velocity.
\par
The Fe\,\textsc{xxv}\,He$\alpha$ complex consists of several prominent emission lines, including $z$, $y$, $x$, and $w$ lines. Resonance scattering is expected to influence the flux of the strongest line, i.e., the $w$ (resonance) line. As shown in Fig.~\ref{fig:rsl_spec} (right), 
the model (blue/red line) appears to underpredict the various emission lines in the Fe\,\textsc{xxv}\,He$\alpha$ complex, except for the $w$ line, which may result from accommodating this brightest line. To test the impact of the effect on our velocity measurements, we performed spectral fittings where this line is excluded, in broad and narrow energy bands. The derived bulk velocities are consistent within the $1.3(1.6)\sigma$ statistical uncertainties for the full (narrow) band fit, and the turbulent velocities are within $0.9(1.4)\sigma$ within the statistical uncertainties to the upper limits estimated from the $w$ line-excluded fits (see Fig.~\ref{fig:spec-tests}). This suggests that resonant scattering has only a minor impact on our results.
\par
To test the influence of a multitemperature structure, we added an additional \texttt{bapec} component (2T). While the Bayesian information criterion test \citep[BIC;][]{BIC} shows that there is no significant improvement in the fit when adding the second component, we report that the LOS bulk velocity resulting from various 2T-fits (e.g., all source parameters from both components are untied and thawed, $\sigma_{\rm v}$ linked, $\sigma_{\rm v}$ and $Z$ linked, and $\sigma_{\rm v}$, $Z$, and $z$ linked) are consistent with the default 1T-fit listed in Table~\ref{tab:results_rsl}.
\par
We assessed the systematic uncertainties originating from the Resolve energy scale (gain) and line spread function. The current Resolve energy-scale accuracy in the $5.4-9.0\,{\rm keV}$ energy band is $\pm0.3\,{\rm eV}$. From the Resolve Energy Scale report\footnote{\url{https://heasarc.gsfc.nasa.gov/docs/xrism/analysis/gainreports/index.html}} of our observation, the worst line shift inferred from the calibration pixel is reported to be $0.11\,{\rm eV}$. Adding these two terms to the quadrature results in a total energy scale uncertainty of $0.32\,{\rm eV}$, which translates to a systematic uncertainty in the LOS bulk velocity of $\sim\!16\,\mathrm{km\,s}^{-1}$ at $6\,{\rm keV}$. Furthermore, the instrumental line spread uncertainty is $\sim\!0.13\,{\rm eV}$ FWHM at $6.0\,{\rm keV}$. This corresponds to a Gaussian width of $\sigma_{\rm E}\approx0.055\,{\rm eV}$ and translates into a systematic uncertainty in the one-dimensional turbulent velocity of $\sim\!3\,\mathrm{km\,s}^{-1}$. Both systematic uncertainties are smaller than the statistical uncertainties of the respective velocity measurements, and therefore do not affect our conclusions.
\par
Lastly, we investigate the impact of the background models used in the Resolve spectral analysis. As shown in Fig.~\ref{fig:rsl_spec}, the CXB (magenta line) is significantly lower than the source spectrum. Even by increasing the CXB by a factor of two, no significant impact is observed on any of the fitted parameters. The used NXB model (cyan line) was also probed by adjusting the normalization by $\pm20\%$. The resulting ICM parameters from these tests are consistent with those from the default fitting.
\par
Therefore, the results of the default spectral analysis (using the unbinned spectrum, in the $2.0-10.0\,\mathrm{keV}$ band, using a single temperature model, and the best-fit CXB model from the Xtend analysis) will be reported and discussed in the following sections. The cluster properties from default Resolve analysis and the Resolve spectrum fitted with the default model are presented in Table~\ref{tab:results_rsl} and Fig.~\ref{fig:rsl_spec} (left), respectively.

\section{Results}\label{sec:results}
\subsection{Galaxy redshift analysis}\label{sec:ned}
\begin{figure}
\centering
    \includegraphics[width=\columnwidth]{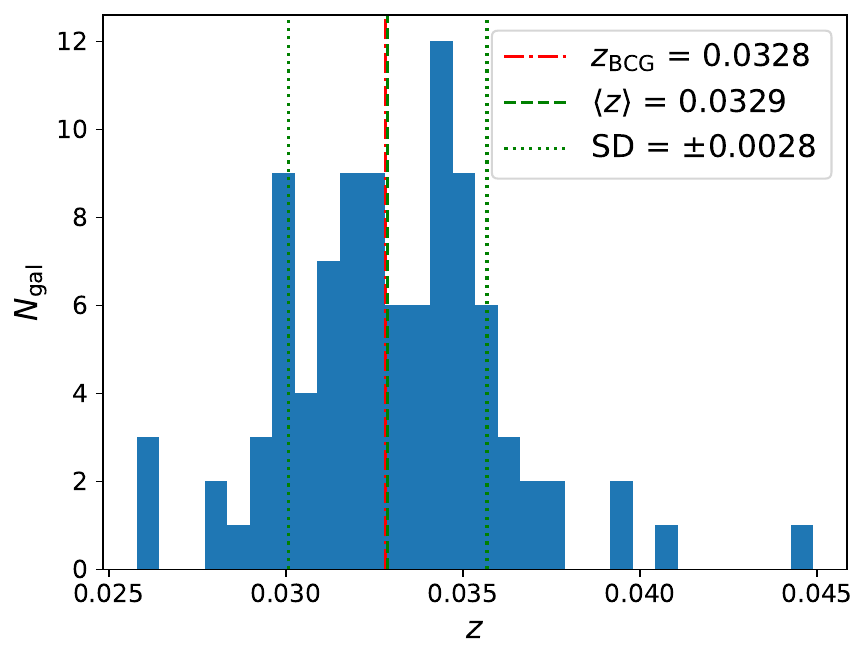}
\caption{Spectroscopic and heliocentric galaxy redshift distribution within $R_{500}$ of A496 compiled from NED. The green dashed and dotted vertical lines mark the mean and the standard deviation of the distribution, while the red dashed-dotted line indicates the redshift of the BCG.}
\label{fig:z-hist}
\end{figure}

To compare the gas motion measured by Resolve with the galaxy component of the A496 cluster, an optical galaxy redshift analysis was carried out. Spectroscopic and heliocentric galaxy redshifts within $R_{500}$ were compiled from the NASA/IPAC Extragalactic Database (NED). The distribution of the 96 galaxies within the boundary is presented in Fig.~\ref{fig:z-hist}. The redshift of the BCG, $z_\mathrm{BCG}=0.03281\pm0.00004$ \citep{Wegner_1999}, is consistent with the mean of the distribution, $\langle z \rangle=0.0329\pm0.0003$, suggesting that the BCG is at rest with respect to the cluster systemic velocity. The standard deviation is estimated to be 0.0028, corresponding to a LOS velocity dispersion of $\sigma_z\approx813\,\mathrm{km\,s^{-1}}$ with respect to the BCG redshift.

\subsection{Spectral analysis}\label{sec:spectro_results}

\begin{table*}
\centering
\caption{Resolve A496 best-fit parameters in the $2.0-10.0\,\mathrm{keV}$ band.}
\resizebox{0.8\textwidth}{!}
{\begin{tabular}{c c c c c c c}
\hline
\hline
$k_\mathrm{B}T$ & $Z$ & $z^\dagger$ & $v_\mathrm{bulk}^{\dagger \ddagger}$ & $\sigma_{\rm v}$ & norm & C-stat/dof \\
$(\mathrm{keV})$ & $(Z_\odot)$ &  & $(\mathrm{km\,s}^{-1})$ & $(\mathrm{km\,s}^{-1})$ & $(10^{-2}\,\mathrm{cm}^{-5})$ & \\
\hline \\[-1.7ex]
$3.23_{-0.08}^{+0.09}$ & $0.664_{-0.039}^{+0.048}$ & $0.03257_{-0.00003}^{+0.00005}$ & $-69_{-20}^{+25}$ & $78_{-16}^{+18}$ & $4.10_{-0.15}^{+0.14}$ & $11569 / 15994$\\[3pt]
\hline
\multicolumn{7}{l}{\footnotesize $^\dagger$ barycentric corrected, $^\ddagger$the errors include the uncertainties from $z_{\rm BCG}$.}\\
\hline
\hline
\end{tabular}}
\label{tab:results_rsl}
\end{table*}

The Resolve spectrum of the A496 cluster and its fitted model in the full $2.0-10.0\,\mathrm{keV}$ band are presented in the left panel of Fig.~\ref{fig:rsl_spec}, revealing a prominent Fe\,\textsc{xxv}\,He$\alpha$ line complex. The right panel shows a similar plot, limited to the narrow band $6.2-6.9\,\mathrm{keV}$ to better illustrate the Fe\,\textsc{xxv}\,He$\alpha$ and Fe\,\textsc{xxvi}\,Ly$\alpha$ line complexes. In both plots, the red and blue lines indicate the total model and the ICM component, respectively. Compared with the source spectrum, the total background, including the NXB (cyan line) and the CXB components (magenta line), appears negligible.
\par
We report an average ICM temperature of $3.23_{-0.08}^{+0.09}\,\mathrm{keV}$, which is in good agreement with the spatially resolved \xmm measurements by \cite{Ghizzardi_2014}. A metallicity value of $0.664_{-0.0386}^{+0.0480}Z_\odot$ is measured, which is also consistent with the reported \xmm core metallicity of A496 in \cite{Lovisari_2019}. A barycentric correction accounting for the spacecraft motion relative to the Solar System barycenter is applied to the obtained redshift, and subsequently, the LOS bulk velocity. The correction at the time of our observation is approximately $-21.4\,\mathrm{km\,s}^{-1}$. The resulting redshift with the barycentric correction is $z=0.03257_{-0.00003}^{+0.00005}$. The LOS bulk velocity is calculated with respect to the BCG redshift by 
\begin{equation}
    v_\mathrm{bulk} = \frac{c(z - z_\mathrm{BCG})}{1 +  z_\mathrm{BCG}},
\end{equation}
which yields $-69_{-20}^{+25}\,\mathrm{km\,s}^{-1}$, where the uncertainty also includes the statistical error from the BCG redshift measurement.
\par
The measured LOS velocity dispersion is $\sigma_{\rm v}=78_{-16}^{+18}\,\mathrm{km\,s}^{-1}$. Assuming that velocity dispersion is due to isotropic turbulence, the 3D Mach number can be derived by

\begin{equation}
    \mathcal{M}_{\rm 3D} = \frac{\sqrt{3}\sigma_{\rm v}}{c_{\rm s}},
\label{eq:m3d}
\end{equation}
where the denominator is the sound speed and is given by

\begin{equation}
c_{\rm s} = \sqrt{\frac{\gamma k_{\rm B}T}{\mu m_{\rm p}}},
\end{equation}
where $\gamma=5/3$ is the adiabatic index and $\mu=0.61$ is the mean molecular weight in a fully ionized plasma. Using the best-fit temperature measured by Resolve (Table~\ref{tab:results_rsl}), we estimated the sound speed at the core of A496 to be $c_{\rm s}=919_{-12}^{+13}\,\mathrm{km\,s}^{-1}$, thus gives $\mathcal{M}_{\rm 3D}=0.15_{-0.03}^{+0.04}$, indicating a subsonic turbulence motion. The non-thermal pressure fraction (accounting only the turbulence motion) can be calculated using the equation from \cite{Eckert_2019},
\begin{equation}
    \frac{P_{\rm NT}}{P_{\rm Tot}} = \frac{\mathcal{M}_{\rm 3D}^2}{\mathcal{M}_{\rm 3D}^2 + 3/\gamma},
    \label{eq:pnt}
\end{equation}
which from Resolve measurement of A496 is $1.2_{-0.5}^{+0.6}\%$.

\section{Discussion}\label{sec:discussion}

\begin{figure*}
    \centering
    \includegraphics[width=0.49\textwidth]{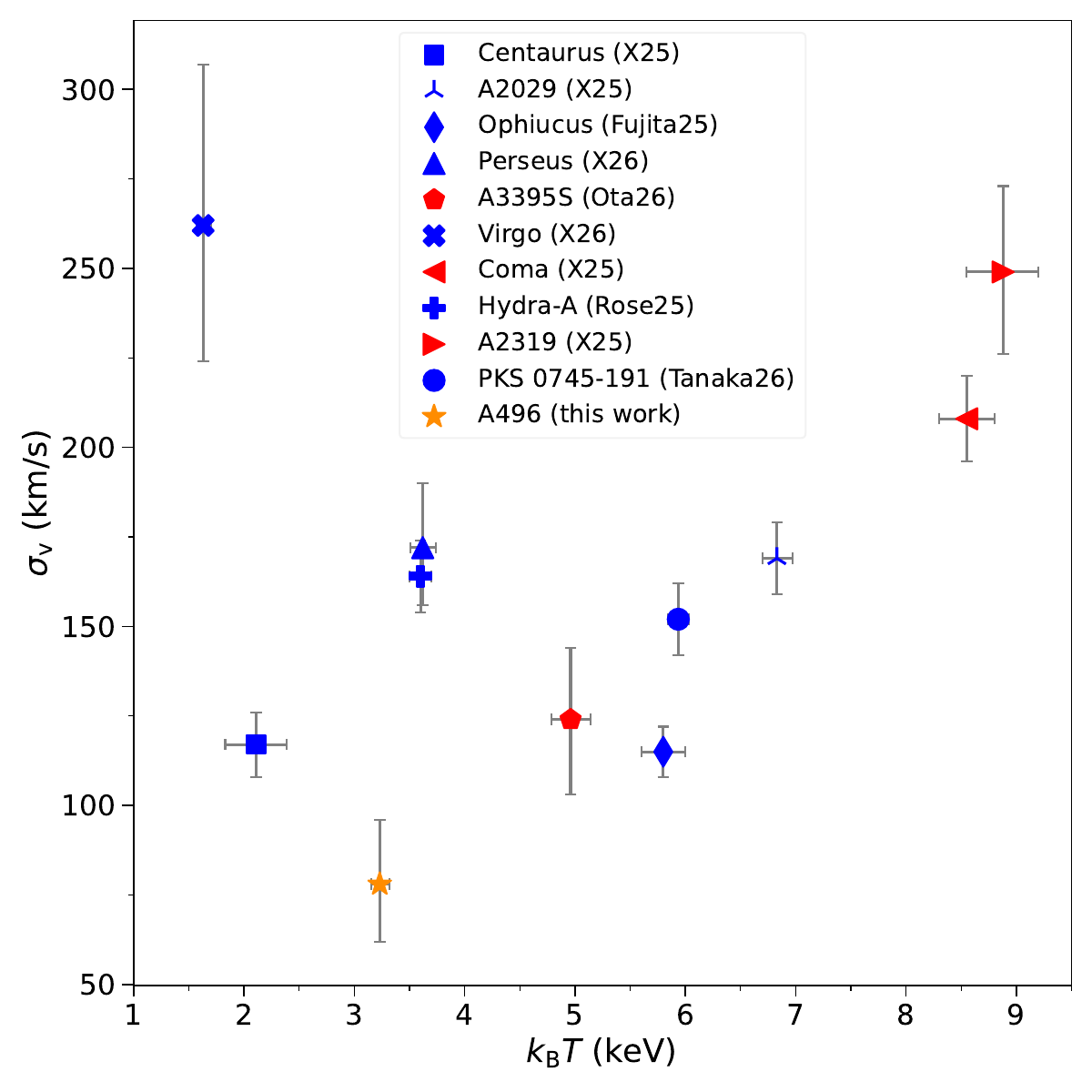}
    \includegraphics[width=0.49\textwidth]{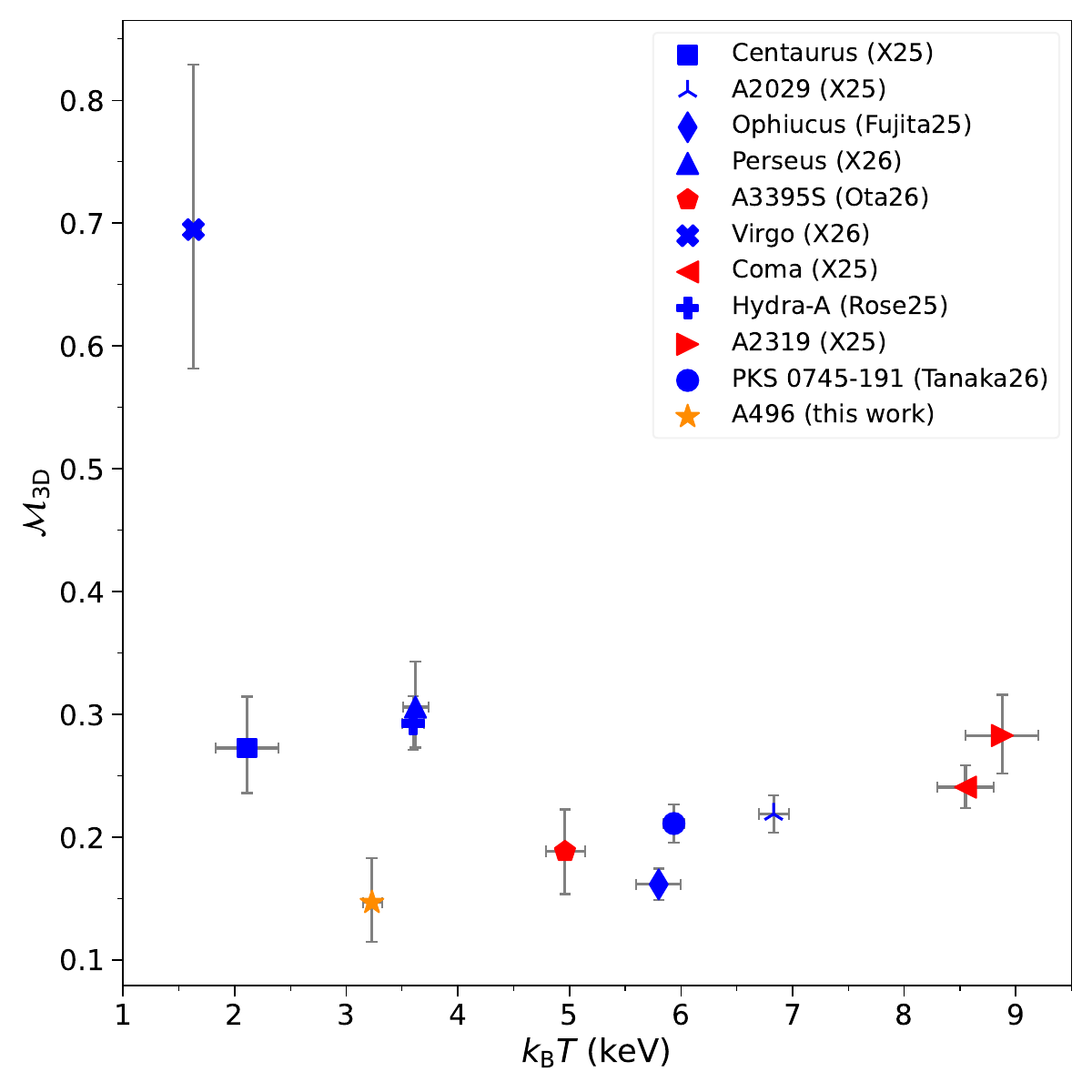}
    \caption{Resolve cluster core kinematics. In each plot, A496 data point from this work is denoted by the orange star. Cool core clusters are plotted in blue (except A496), while non-cool core are in red. \textit{Left}: $\sigma_{\rm v}-k_{\rm B}T$. \textit{Right}: $\mathcal{M}_{\rm 3D}-k_{\rm B}T$}
    \label{fig:xrism_parspaces}
\end{figure*}

\subsection{Bulk Velocity}
Resolve reveals a dynamically quiescent state of the ICM in the A496 core. Despite some sloshing-induced features, such as multiple cold fronts and an enhanced spiral pattern identified by previous studies using \xmm and \chandra observations \citep[e.g.,][]{Tanaka_2006, Dupke_2007, Lagana_2010, Ghizzardi_2014}, which are indicative of coherent gas motion \citep{Markevitch_2007}, the line-of-sight bulk velocity of ICM relative to the motion of the BCG is moderate with $v_{\rm bulk}=-69_{-20}^{+25}\,{\rm km\,s^{-1}}$.
In dedicated A496 simulations, these features were predicted to be triggered by an off-axis minor merger \citep{Roediger_2012}. The Simulating the LOcal Web (SLOW) constrained Universe simulation \citep{SLOW_Sorce, Dolag_2023, SLOW_Hernandez, SLOW_Seidel, Groth_2026} further supports a merger configuration consistent with that proposed by \cite{Roediger_2012}.
\par
We further compare the observations to the non-radiative \textsc{LowerDecks} zoom-in simulation \citep{Seidel_2026} based on SLOW with the same setup as described by \citet{Groth_2026}, using emissivity-weighted projected velocity maps similar to \citet{Vazza&Brunetti2026}. As uncertainties can arise due to timing differences of the simulation compared to the observed cluster, we analyze the redshift evolution close to $z=0$, where we have simulation output available. The gas bulk velocity in the center relative to the BCG is estimated to be $\approx 0\,{\rm km\,s^{-1}}$ at $z=0.001$, and reaches $\mathcal{O}(-100)\,{\rm km\,s^{-1}}$ at $z=0$, fully consistent with XRISM measurements. This indicates that no AGN feedback is required, consistent with the cool-core nature of the cluster.
\par
Compared to other bulk velocity measurements of CC clusters by XRISM, our value lies between the most relaxed cluster known, A2029 \citep{XRISM_A2029, XRISM_A2029_Sarkar} and the Centaurus cluster \citep{XRISM_Cent}. While there appears to be no trend among the CC clusters, this may indicate that the clusters are at different stages of sloshing, as also shown by the bulk velocity evolution in the SLOW simulations. For example, it is suggested that the observed low bulk velocity of the Ophiuchus cluster, $\lvert v_{\rm bulk} \rvert = 8\pm7\,{\rm km\,s^{-1}}$, might be because its sloshing core is approaching a turning point \citep{Fujita_2025}.

\subsection{Turbulence and central AGN feedback}\label{sec:discussion_turb}

\begin{figure*}
    \centering
    \includegraphics[width=0.49\textwidth]{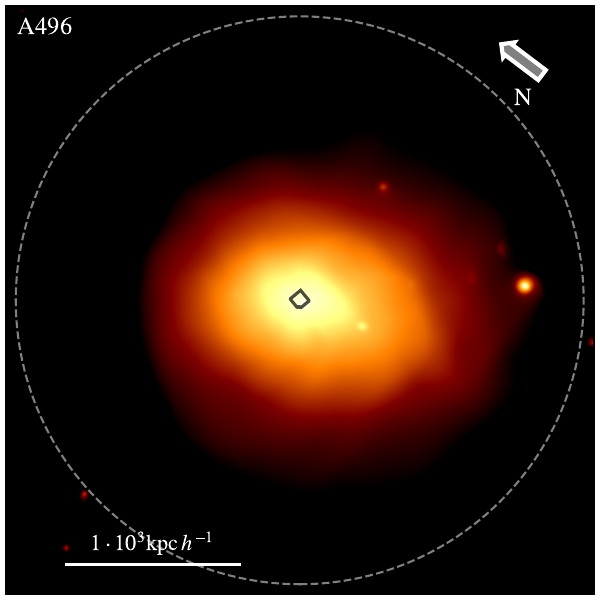}
    \includegraphics[width=0.49\textwidth, trim=0.0cm 0.2cm 0.0cm 0.0cm, clip]{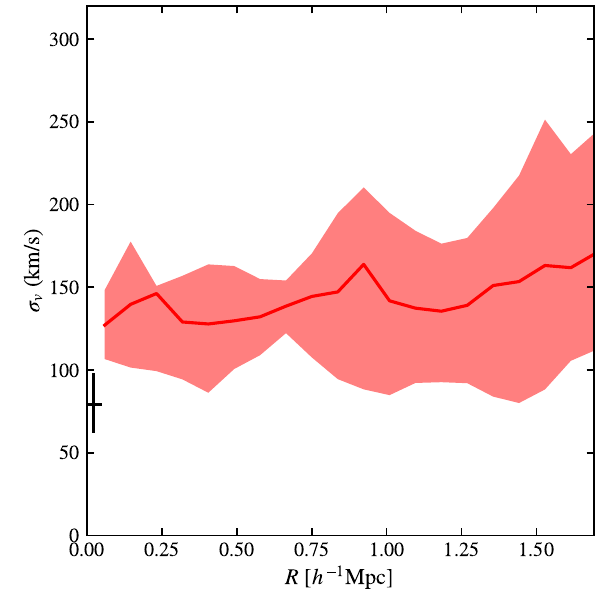}
    \caption{The SLOW constrained Universe simulation of A496. \textit{Left}: X-ray luminosity in the $0.5-7.0\,{\rm keV}$ in a 5\,Mpc regions taken at a snapshot that corresponds to $z=0.001$ The dashed circle indicates the virial radius in the simulation. The Resolve FoV of our existing central observation is denoted by a black box in the center. Within this FoV, the simulation estimated an emission-weighted $\sigma_{\rm 1D} = 125\pm25\,{\rm km\,s^{-1}}$. \textit{Right}: One-dimensional turbulent velocity profiles from the SLOW simulation of A496. The black cross indicates the Resolve measurement.}
    \label{fig:SLOW_A496}
\end{figure*}

As illustrated in Fig.~\ref{fig:xrism_parspaces} (right panel), the ICM velocity dispersion in the core of A496, $\sigma_{\rm v}=78_{-16}^{+18}\,{\rm km\,s^{-1}}$, is the lowest among some of the available XRISM cluster core measurements: Virgo \citep{XRISM_Virgo}, Centaurus \citep{XRISM_Cent}, Perseus \citep{XRISM_Perseus}, Hydra-A \citep{XRISM_HydraA}, A3395S \citep{Ota_2026}, Ophiuchus \citep{Fujita_2025}, A2029 \citep{XRISM_A2029}, Coma \citep{XRISM_Coma}, A2319 \citep{XRISM_A2319}, PKS 0745-191 \citep{XRISM_PKS}. However, when comparing the velocity dispersions among these clusters, it is important to note that the observed spectra are dominated by regions along the LOS with higher emissivity. \cite{Zhuravleva_2012} introduced the effective length $l_{\rm eff}$, defined as a characteristic scale at a projected distance $R$ within which $50\%$ of the cumulative X-ray flux along the full LOS is enclosed. Since typically X-ray emissivity peaks towards cluster centers, this means that $l_{\rm eff}$ is shorter at the center and increases with increasing $R$. For instance, using $\beta$-model parameters of A496 obtained using \xmm data listed in \cite{Lagana_2008} to describe the ICM distribution ($\beta=0.410\pm0.015$ and $r_{\rm c}=15.5\pm0.2\,{\rm kpc}$), we estimated $l_{\rm eff}$ to be $21-24\,{\rm kpc}$ at the center and $84-93\,{\rm kpc}$ at $R=59\,{\rm kpc}$ ($\approx1.5'$, which is half width of Resolve FoV). For further discussion, we adopt the mean value, $l_{\rm eff}\sim57\,{\rm kpc}$, which is comparable to the effective length scale of the Perseus cluster \cite{XRISM_Perseus}. Given that the velocity dispersions from the different clusters originate from regions with varying $l_{\rm eff}$, a direct comparison of this parameter is not suitable. We show the 3D mach number (Eq.~\ref{eq:m3d}) as a function of temperature in Fig.~\ref{fig:xrism_parspaces} (right). In the low temperature regime ($k_{\rm B}T\leq 4\,{\rm keV}$), A496 exhibits the lowest $\mathcal{M}_{\rm 3D}$. Compared to Centaurus and Perseus, A496 may represent one of the most quiescent sloshing cores observed so far.
\par
Furthermore, the fraction of non-thermal pressure contributed by turbulence to total pressure in the A496 cluster derived from Resolve measurement is only $1.2_{-0.5}^{+0.6}\%$ (see Sect.~\ref{sec:spectro_results}), namely, the core of A496 is dominated by thermal pressure and the central AGN appears to have a negligible contribution to the kinematics of the surrounding ICM. This value is lower than the predictions. \cite{XRISM_sigma-sim} found that the kinetic-to-thermal pressure ratios of the different simulations are higher than the observed values from the cores of CC clusters (except for the Virgo cluster). From the simulations, the median of the level of kinetic pressure support varies between $4.0\%$ and $7.0\%$, while the median of the values measured by Resolve is $2.2_{-1.0}^{+2.0}\%$ \citep{XRISM_sigma-sim}.
\par
From the formation history of A496 alone (i.e., without feedback), the SLOW simulation predicts a higher 1D emission-weighted velocity dispersion of $125\pm25\,{\rm km\,s^{-1}}$ specifically in this Resolve FoV (Fig.~\ref{fig:SLOW_A496}). This estimate was calculated at $z=0.001$. The value does not change from $z=0.001$ to $z=0$, as there has been no recent major merger event, indicating that this result is unaffected by timing uncertainties. While the velocity dispersion is slightly above the observed value, it is a better match than previous statistical comparisons. Selection effects are thus important to explain these differences \citep{Groth_2026}. In addition, SLOW predicts an increasing velocity dispersion gradient to the outskirts, consistent with predictions from sloshing motions \citep{XRISM_Perseus_sim}. Nevertheless, the overestimation in simulations is also presented in \cite{XRISM_sigma-sim}. The authors compare the available Resolve measurements with three cosmological simulation suites of galaxy clusters that incorporate different baryonic physics models and all include key physical processes relevant to the formation and evolution of galaxies. While the predicted velocity dispersions in the central pointings of the CC clusters are in agreement among the different simulation suites, the observed values from Resolve are systematically below the simulation medians by a factor of $1.5-1.7$ \citep{XRISM_sigma-sim}. They suggest that the discrepancy between the predicted and observed velocity dispersion in the CC clusters might be due to incomplete modeling of ICM physics in simulations; for instance, AGN feedback may be too ejective.
\par
The idealized numerical simulations aimed at disentangling the roles of merger-induced sloshing and AGN feedback in Perseus-like clusters show that these two processes, separately and together, influence velocity dispersion profiles in clusters, especially in the innermost regions \citep{XRISM_Perseus_sim}. At $r\lesssim50\,{\rm kpc}$, the velocity dispersion values vary between $\sim\!50\,{\rm km\,s^{-1}}$ for the sloshing-only model and $\sim\!200\,{\rm km\,s^{-1}}$ for AGN-only model ($P_{\rm jet}\sim1\times10^{45}\,{\rm erg\,s^{-1}}$). Our value lies between the sloshing-only and sloshing+AGN feedback of $3\times10^{44}\,{\rm erg\,s^{-1}}$ models. To compare with their predictions, we estimated the mechanical feedback from the most recent activity of the A496 central radio source. We note that a visual inspection of the \chandra image reveals no X-ray AGN corresponding to the central radio source. The total radio flux density of the core and the jets from the VLA at 5\,GHz is $S_{\rm 5\,GHz}=57\pm6\,{\rm mJy}$ \citep{Ubertosi_2024}. This innermost radio component has a largest linear size (LLS) of about $10\,{\rm kpc}$. Using the flux to luminosity conversions \citep[Eq.\,2 and 3 in][]{Veronica_2026}, adopting $\alpha=0.7$ from \cite{Ubertosi_2024}, and the mechanical AGN power equation from \cite{Heckman_2014}, we estimated a value of $(1.27\pm0.12)\times10^{43}\,{\rm erg\,s^{-1}}$. This outcome seems to align with the prediction from \cite{XRISM_Perseus_sim}. Additionally, we estimated the bolometric luminosity from Resolve FoV, which is $(1.57\pm0.02)\times10^{44}\,{\rm erg\,s^{-1}}$. The half-width of Resolve FoV corresponds to $0.07R_{500}$ (Table~\ref{tab:A496info}), which is of the same order of the cooling radius determined by \cite{Hudson_2010} using the CC clusters in the HIghest X-ray FLUx Galaxy Cluster Sample \citep[\textit{HIFLUGCS};][]{RB_2002}, which is $R_{\rm cool}=0.08R_{500}$. Comparing the radio mechanical power and X-ray cooling luminosity, the contribution of the central AGN to the ICM heating through mechanical feedback is $7-9\%$, which is consistent with the efficiency derived from the relation between AGN feedback and cooling luminosity in the eRASS1/ASKAP CC cluster sample in \cite{Veronica_2026}.
\par
Furthermore, we conducted a first-order two-region spectral analysis (no PSF-mixing modeling), namely, the inner $4\times4$ pixels and outer 18 pixels, to investigate for any hint of spatial dependence in the velocity measurements. We found that the turbulent velocities of the inner and outer regions are consistent with each other within their $1\sigma$ statistical uncertainties, and their bulk velocities agree well within their $1.5\sigma$ uncertainties (see Fig~\ref{fig:spec-tests}). These results suggest little to no velocity gradient or enhanced turbulence near the central AGN.
Since the LLS of the central component is smaller than $l_{\rm eff}$ (assuming it lies in the plane of the sky), in addition to it being a relatively weak central AGN, turbulence driven by the current AGN activity may be diluted in the observed LOS emissivity-weighted Resolve spectrum, hence, might further explain the relatively low turbulent velocity.

\subsection{Multiphase structures in the core}\label{sec:discussion_multiphase}
\begin{figure}
    \centering
    \includegraphics[width=\columnwidth]{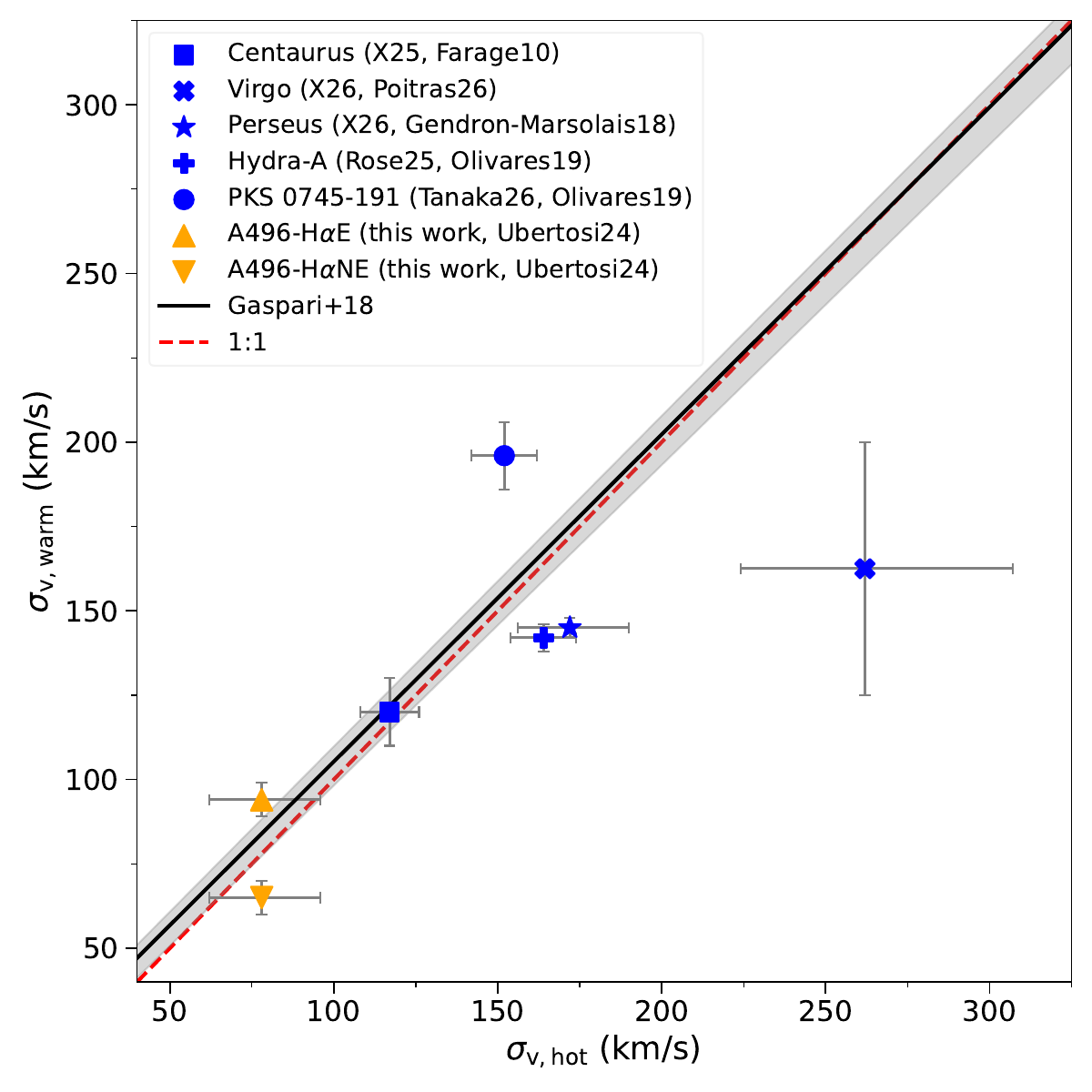}
    \caption{Warm against hot-gas velocity dispersion of some Resolve CC sample. The $\sigma_{\rm v.\,hot}$ are Resolve measurements from recent XRISM literature (see Sect.~\ref{sec:discussion_turb}), while the $\sigma_{\rm v,\,warm}$ values are obtained from cluster core filamentary warm gas structure measurements from various literature (see Sect.~\ref{sec:discussion_multiphase}). The solid black line and the gray shaded area are the best-fit correlation and its $1\sigma$ confidence interval from \cite{Gaspari_2018}. The red-dashed line indicates the 1-to-1 line.}
    \label{fig:sigma_hot-warm}
\end{figure}

In the scenario of a top-down multiphase condensation cascade, warm and cold filaments and clouds form as the hot plasma condenses. A fraction of this condensed gas is accreted onto the central supermassive black hole (SMBH), triggering self-regulating AGN feedback \citep[cold chaotic accretion (CCA);][]{Gaspari_2013}. The condensed structures may retain some properties of the hot plasma \citep{Gaspari_2018}, for example, according to their model, there is a tight correlation in the LOS velocity dispersion between hot and warm gas \citep[see Fig.~1 of][their best-fit correlation is also shown as the solid black line with the gray shaded area in Fig.~\ref{fig:sigma_hot-warm}]{Gaspari_2018}, as well as the positive linear correlation in their metallicity \citep[see Fig.~3 of][]{Olivares_2026}. A strong correlation between the morphology of warm and hot gases has also been reported by \cite{McDonald_2010} in the sample study of H$\alpha$ filaments in CC clusters. \cite{Olivares_2025} further found a tight positive correlation between the X-ray and the H$\alpha$ surface brightness of the filaments in the cores of seven CC clusters.
\par
To investigate the multiphase structures in the core of A496, we compare our results with the warm gas structure studied in \cite{Ubertosi_2024}. These authors utilized optical observations of A496 with the MUSE integral-field spectrograph mounted on the Very Large Telescope to derive the H$\alpha$-line intensity and kinematics in the cluster core. The map reveals an extended warm ($10^4\,{\rm K}$) gas nebula surrounding the BCG. The structure is composed of a bright central region, a $\sim\!10\,{\rm kpc}$ filament extending toward the east, and a slightly longer, fainter filament toward the northeast. The bright region of the warm gas nebula aligns with the core of the radio structure (Sect.~\ref{sec:intro}), and its overall morphology is co-spatial with the regions of enhanced X-ray emission \citep[see Fig.~9b of][]{Ubertosi_2024}. Additionally, they reported that the A496 eastern warm gas filament has a measured velocity dispersion of $94\pm5\,{\rm km\,s^{-1}}$, while the more extended northeastern filament has a velocity dispersion of $65\pm5\,{\rm km\,s^{-1}}$. These values are in good agreement with our XRISM hot gas velocity dispersion, supporting that at least some warm gas may originate from ICM condensation in the wake of the radio bubble \citep{Ubertosi_2024}.
\par
Moreover, in Fig.~\ref{fig:sigma_hot-warm} we also plotted data points for five CC clusters with central warm filamentary nebulae measurements, including Centaurus \citep{Farage_2010}, Virgo \citep{Poitras_2026}, Perseus \citep{Gendron_2018}, as well as Hydra-A and PKS 0745-191 \citep{Olivares_2019}.
A good agreement with the expected correlation from the model in \cite{Gaspari_2018} is observed in the low $\sigma_{\rm v,\,warm}-\sigma_{\rm v,\,hot}$ regime, while at $\sigma_{\rm v,\,hot}>150\,{\rm km\,s^{-1}}$ a deviation from the correlation arises. In the case of Virgo, if it were located at a higher redshift, for instance, as such, Resolve would only see the average of three central bins ($5-25\,{\rm kpc}$), and then the data point would lie much closer to the 1-to-1 line.
Nonetheless, the outliers still lie within the intrinsic scatter of the simulated sample in \cite{Gaspari_2018}. In summary, the observational data shown in Fig.~\ref{fig:sigma_hot-warm} clearly establish a correlation between the measured central velocity dispersions of the warm and hot gas phases.
A larger sample is required, in particular to investigate the high $\sigma_{\rm v,\,warm}-\sigma_{\rm v,\,hot}$ regime.

\section{Summary and conclusions}\label{sec:conclude}
Gas motions provide insight into the dynamical history and physical processes within galaxy clusters. Large-scale motions are typically driven by structure formation, such as mergers and accretion, while small-scale motions are governed by processes, such as AGN feedback, gas sloshing, and instabilities. In this work, we probed the dynamics and kinematics of the ICM in the core of A496 using XRISM/Resolve data. We discussed the ICM kinematics and compared it with other Resolve cluster core measurements. We also compared our results with simulations and multiwavelength observations. Our findings are,
\begin{itemize}
    \item Optical redshift analysis confirmed that the BCG \citep[$z_{\rm BCG}=0.03281\pm0.00004$;][]{Wegner_1999} is at rest with respect to the cluster systemic velocity ($\langle z \rangle=0.0329\pm0.0003$).
    \item Despite multiple previously detected sloshing-induced features, such as cold fronts and an enhanced spiral pattern, and the presence of a (weak) central radio source, Resolve observation reveals that the core of A496 is dynamically quiescent with a LOS bulk velocity of $v_{\rm bulk}=-69_{-20}^{+25}\,{\rm km\,s^{-1}}$ with respect to the BCG and a turbulent velocity of $\sigma_{\rm v}=78_{-16}^{+18}\,{\rm km\,s^{-1}}$. The $\sigma_{\rm v}$ value is one of the lowest measured in the core of any cluster by the instrument considered in this work. We note that a more quantitative comparison should account for the physical scale sampled by Resolve. Assuming isotropic turbulence, this value corresponds to a subsonic 3D Mach number of $0.15_{-0.03}^{+0.04}$ and a non-thermal pressure fraction of $1.2_{-0.5}^{+0.6}\%$.
    \item Previous studies showed that A496's H$\alpha$ filament in the core is spatially correlated with an ICM surface brightness enhancement. In this work, we show that the ICM 1D velocity dispersion is in good agreement with the velocity dispersion of the H$\alpha$ filament in the core. Moreover, by adding literature values of five clusters, we establish a correlation between the central warm and hot gas velocity dispersions, consistent with $\frac{\sigma_{\rm v,hot}}{\sigma_{\rm v,warm}}=1$. This may support the notion that at least some of this warm gas originates from the condensation of the ICM in the wake of radio bubbles.
    \item We estimated the mechanical AGN feedback from the recent activity of the central radio source and found a level of contribution of $7-9\%$ to ICM heating. Since the largest linear size of the central radio component is smaller than the approximated effective length within the FoV, turbulence driven by the current AGN activity may be diluted in the observed LOS emissivity-weighted spectrum, thereby explaining the relatively low turbulent velocity.
    \item We compared our velocity measurements with the Simulating the LOcal Web (SLOW) constrained Universe simulation. The 1D LOS bulk velocity from SLOW is consistent with the measured value, suggesting that AGN feedback has a negligible contribution. The A496 SLOW turbulent velocity, as in other reported Resolve$-$simulation comparisons, is higher, but remains within $1.5\sigma$ uncertainty.
\end{itemize}

\begin{acknowledgements}
      We thank Dominik Kox and Lina Gerlach for their early work on the A496 cluster using eRASS data.
      AV acknowledges support from the German Federal Ministry of Research, Technology and Space (BMFTR) provided through the German Space Agency (DLR) under project 50OR2518.
      Partially funded by the Deutsche Forschungsgemeinschaft (DFG, German Research Foundation) under Germany’s Excellence Strategy – EXC 3037 – 533607693.
      FG, KD and BAS acknowledge support by the COMPLEX project from the European Research Council (ERC) under the European Union’s Horizon 2020 research and innovation program grant agreement ERC-2019-AdG 882679.
      FG and KD acknowledge support by the Deutsche Forschungsgemeinschaft (DFG, German Research Foundation) under Germany’s Excellence Strategy - EXC-2094 - 390783311.
      YZ acknowledges support from the Chinese Scholarship Council (CSC) and the German Academic Exchange Service (DAAD).
      We thank the XRISM project and the mission operations team for the operation of the satellite and their support of the observations.
      This work is based on data from eROSITA, the soft X-ray instrument aboard SRG, a joint Russian-German science mission supported by the Russian Space Agency (Roskosmos), in the interests of the Russian Academy of Sciences represented by its Space Research Institute (IKI), and the Deutsches Zentrum für Luft- und Raumfahrt (DLR). The SRG spacecraft was built by Lavochkin Association (NPOL) and its subcontractors, and is operated by NPOL with support from the Max Planck Institute for Extraterrestrial Physics (MPE). The development and construction of the eROSITA X-ray instrument was led by MPE, with contributions from the Dr. Karl Remeis Observatory Bamberg and ECAP (FAU Erlangen-Nuernberg), the University of Hamburg Observatory, the Leibniz Institute for Astrophysics Potsdam (AIP), and the Institute for Astronomy and Astrophysics of the University of Tübingen, with the support of DLR and the Max Planck Society. The Argelander Institute for Astronomy of the University of Bonn and the Ludwig Maximilians Universität Munich also participated in the science preparation for eROSITA. The eROSITA data shown here were processed using the eSASS software system developed by the German eROSITA consortium.
      We thank the staff of the GMRT that made these observations possible. GMRT is run by the National Centre for Radio Astrophysics of the Tata Institute of Fundamental Research.
      The Legacy Surveys consist of three individual and complementary projects: the Dark Energy Camera Legacy Survey (DECaLS; Proposal ID \#2014B-0404; PIs: David Schlegel and Arjun Dey), the Beijing-Arizona Sky Survey (BASS; NOAO Prop. ID \#2015A-0801; PIs: Zhou Xu and Xiaohui Fan), and the Mayall z-band Legacy Survey (MzLS; Prop. ID \#2016A-0453; PI: Arjun Dey). DECaLS, BASS and MzLS together include data obtained, respectively, at the Blanco telescope, Cerro Tololo Inter-American Observatory, NSF’s NOIRLab; the Bok telescope, Steward Observatory, University of Arizona; and the Mayall telescope, Kitt Peak National Observatory, NOIRLab. Pipeline processing and analyses of the data were supported by NOIRLab and the Lawrence Berkeley National Laboratory (LBNL). The Legacy Surveys project is honored to be permitted to conduct astronomical research on Iolkam Du’ag (Kitt Peak), a mountain with particular significance to the Tohono O’odham Nation.
      NOIRLab is operated by the Association of Universities for Research in Astronomy (AURA) under a cooperative agreement with the National Science Foundation. LBNL is managed by the Regents of the University of California under contract to the U.S. Department of Energy.
      This project used data obtained with the Dark Energy Camera (DECam), which was constructed by the Dark Energy Survey (DES) collaboration. Funding for the DES Projects has been provided by the U.S. Department of Energy, the U.S. National Science Foundation, the Ministry of Science and Education of Spain, the Science and Technology Facilities Council of the United Kingdom, the Higher Education Funding Council for England, the National Center for Supercomputing Applications at the University of Illinois at Urbana-Champaign, the Kavli Institute of Cosmological Physics at the University of Chicago, Center for Cosmology and Astro-Particle Physics at the Ohio State University, the Mitchell Institute for Fundamental Physics and Astronomy at Texas A\&M University, Financiadora de Estudos e Projetos, Fundacao Carlos Chagas Filho de Amparo, Financiadora de Estudos e Projetos, Fundacao Carlos Chagas Filho de Amparo a Pesquisa do Estado do Rio de Janeiro, Conselho Nacional de Desenvolvimento Cientifico e Tecnologico and the Ministerio da Ciencia, Tecnologia e Inovacao, the Deutsche Forschungsgemeinschaft and the Collaborating Institutions in the Dark Energy Survey. The Collaborating Institutions are Argonne National Laboratory, the University of California at Santa Cruz, the University of Cambridge, Centro de Investigaciones Energeticas, Medioambientales y Tecnologicas-Madrid, the University of Chicago, University College London, the DES-Brazil Consortium, the University of Edinburgh, the Eidgenossische Technische Hochschule (ETH) Zurich, Fermi National Accelerator Laboratory, the University of Illinois at Urbana-Champaign, the Institut de Ciencies de l’Espai (IEEC/CSIC), the Institut de Fisica d’Altes Energies, Lawrence Berkeley National Laboratory, the Ludwig Maximilians Universitat Munchen and the associated Excellence Cluster Universe, the University of Michigan, NSF’s NOIRLab, the University of Nottingham, the Ohio State University, the University of Pennsylvania, the University of Portsmouth, SLAC National Accelerator Laboratory, Stanford University, the University of Sussex, and Texas A\&M University.
      BASS is a key project of the Telescope Access Program (TAP), which has been funded by the National Astronomical Observatories of China, the Chinese Academy of Sciences (the Strategic Priority Research Program ”The Emergence of Cosmological Structures” Grant \# XDB09000000), and the Special Fund for Astronomy from the Ministry of Finance. The BASS is also supported by the External Cooperation Program of Chinese Academy of Sciences (Grant \# 114A11KYSB20160057), and Chinese National Natural Science Foundation (Grant \# 12120101003, \# 11433005).
      The Legacy Survey team makes use of data products from the Near-Earth Object Wide-field Infrared Survey Explorer (NEOWISE), which is a project of the Jet Propulsion Laboratory/California Institute of Technology. NEOWISE is funded by the National Aeronautics and Space Administration.
      The Legacy Surveys imaging of the DESI footprint is supported by the Director, Office of Science, Office of High Energy Physics of the U.S. Department of Energy under Contract No. DE-AC02-05CH1123, by the National Energy Research Scientific Computing Center, a DOE Office of Science User Facility under the same contract; and by the U.S. National Science Foundation, Division of Astronomical Sciences under Contract No. AST-0950945 to NOAO.
      This research has made use of the NASA/IPAC Extragalactic Database (NED), which is funded by the National Aeronautics and Space Administration and operated by the California Institute of Technology.
      This work made use of the Python packages \texttt{NumPy}, \texttt{SciPy}, \texttt{uncertainties}, and \texttt{Astropy}. ChatGPT has been used for code completion and generation of short Python scripts, e.g., for plotting.
\end{acknowledgements}

%
%
\bibliographystyle{aa}
\bibliography{list_bib}

\onecolumn
\begin{appendix}
\section{Resolve spectral analysis systematic tests}
\begin{figure*}[!ht]
\centering
    \includegraphics[width=\columnwidth]{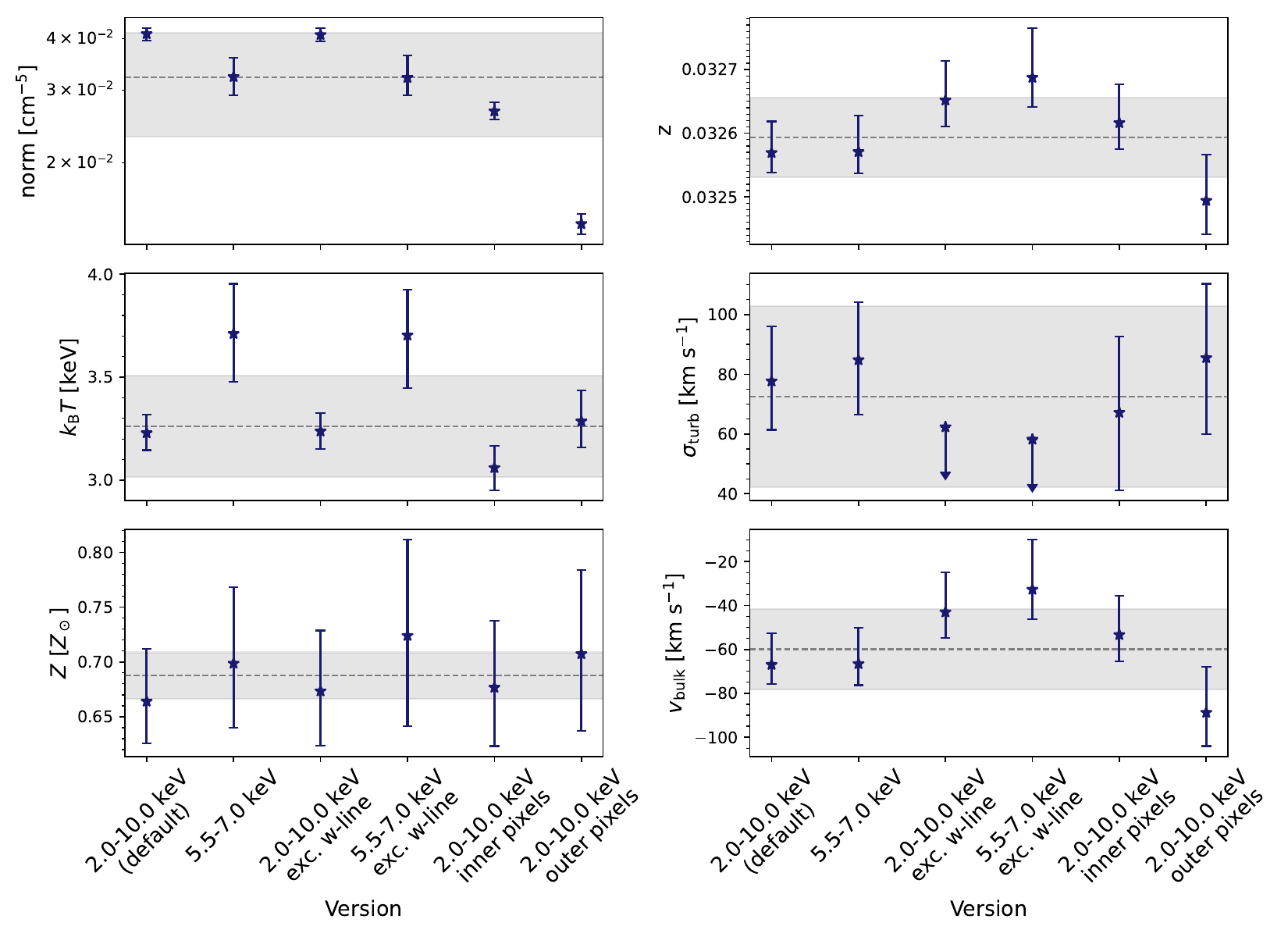}
\caption{ICM parameters from Resolve spectral analysis tests. Note on the labels on x-axes: The first line indicates the energy band in which the fit was performed, "exc. w-line" is spectral fitting without the $w$ (resonance) line, "inner(outer) pixels" is spectral fitting utilizing the inner(outer) $4\times4$(18) pixels. Downward arrows indicate upper limit measurements. The horizontal dashed lines and the gray shaded areas show the median and the standard deviation of the best-fit values from the different tests. For a more detailed description of the tests, see Sect.~\ref{sec:systematics} and \ref{sec:discussion_turb}.}
\label{fig:spec-tests}
\end{figure*}

\end{appendix}

\end{document}